\documentclass[12pt]{amsart}
\usepackage{amsfonts,amssymb,amsthm}
\usepackage[mathscr]{eucal}
\usepackage{hyperref}
\usepackage{times}

\advance\textheight\topskip
\renewcommand{\tilde}{\widetilde}
\renewcommand{\hat}{\widehat}
\newtheorem{theorem}{Theorem}[section]
\newtheorem{prop}{Proposition}[section]

\newtheorem{definition}[prop]{Definition}

\renewcommand{\simeq}{\cong}

\newcommand{\bref}[1]{\textbf{\ref{#1}}}

\newcommand{\gh}[1]{\mathrm{gh}(#1)}

\def\P{{ \bf P}}
\def\G{{ \bf G}}

\newcommand{\aA}{\mathfrak{A}}
\newcommand{\qA}{\hat{\mathfrak{A}}}

\newcommand{\map}{\,\mathrm{:}\,}

\newcommand{\dd}{\partial}
\renewcommand{\d}{\partial}

\newcommand{\tensor}{\otimes}

\renewcommand{\geq}{\,{\geqslant}\,}
\renewcommand{\leq}{\,{\leqslant}\,}

\newcommand{\binner}[2]{%
  {\langle}\kern-4.15pt{\langle}#1{,}\,#2{\rangle}\kern-4.15pt{\rangle}}
\newcommand{\commut}[2]{[#1{,}\,#2]}

\newcommand{\pb}[2]{\left\{{}#1{},{}#2{}\right\}}

\newcommand{\half}{\mathchoice{%
    \ffrac{1}{2}}{\frac{1}{2}}{\frac{1}{2}}{\frac{1}{2}}}

\newcommand{\ffrac}[2]{\raisebox{.5pt}%
  {\footnotesize$\displaystyle\frac{#1}{#2}$}\kern1pt}

\newcommand{\brst}{\mathsf{\Omega}}

\newcommand{\st}[2]{\overset{#1}{#2}}
\newcommand{\dl}[1]{\mathchoice{\ffrac{\dd}{\dd #1}}{\frac{\dd}{\dd
      #1}}{\ffrac{\dd}{\dd #1}}{\ffrac{\dd}{\dd #1}}}

\newcommand{\dr}[1]{\mathchoice{\ffrac{\overset{\leftarrow}{\dd}}{\dd
      #1}}{\frac{\overset{\leftarrow}{\dd}}{\dd #1}}{\ffrac{\overset{\leftarrow}{\dd}}{\dd
      #1}}{\ffrac{\overset{\leftarrow}{\dd}}{\dd #1}}}

\newcommand{\manifold}[1]{\mathscr{#1}}

\newcommand{\manM}{\manifold{M}}

\def\cC{\mathcal{C}}

\def\cE{\mathcal{E}}

\def\cG{\mathcal{G}}

\def\cP{\mathcal{P}}

\def\cW{\mathcal{W}}

\numberwithin{equation}{section} \makeatletter
\@addtoreset{equation}{section}

\def\@secnumfont{\bfseries}
\def\subsubsection{\@startsection{subsubsection}{3}%
  \z@{.5\linespacing\@plus.7\linespacing}{-.5em}%
  {\normalfont\bfseries}}
\def\paragraph{\@startsection{paragraph}{4}%
  \z@\z@{-\fontdimen2\font}%
  \normalfont\bfseries}
\def\subparagraph{\@startsection{subparagraph}{5}%
  \z@\z@{-\fontdimen2\font}%
  \normalfont\bfseries}

\makeatother

\def\mR{\mathcal{R}}
\def\mD{\mathcal{D}}

\begin{document}
\pagestyle{myheadings}
\markboth{\textsc{\small Batalin, Grigoriev, and Lyakhovich}}{
  \textsc{\small Non-Abelian Conversion}}
\addtolength{\headsep}{4pt}

\begin{flushright}
FIAN-TD/05-01\\
hep-th/0501097
\end{flushright}

\vspace{0.5cm}

\begin{centering}

  \vspace{1cm}

  \textbf{\Large{Non-Abelian Conversion and Quantization of\\[10pt]
 Non-scalar Second-Class Constraints}}

  \vspace{1.5cm}

  {\large I.~Batalin,$^{a}$ \ M.~Grigoriev,$^{a}$ and
    S.~Lyakhovich$^b$}

  \vspace{1cm}

  \begin{minipage}{.8\textwidth}\small
    \mbox{}\kern-4pt$^a$Tamm Theory Department, Lebedev Physics
    Institute, Leninsky prospect 53, 119991 Moscow, Russia\\[6pt]
    \mbox{}\kern-4pt$^b$Tomsk State University, prospect Lenina 36,
    634050 Tomsk, Russia
  \end{minipage}

\end{centering}

\vspace{1.5cm}

\begin{center}
  \begin{minipage}{.95\textwidth}
    \textsc{Abstract}.  We propose a general method for deformation
    quantization of any second-class constrained system on a
    symplectic manifold. The constraints determining an arbitrary
    constraint surface are in general defined only locally and can be
    components of a section of a non-trivial vector bundle over the
    phase-space manifold. The covariance of the construction with
    respect to the change of the constraint basis is provided by
    introducing a connection in the ``constraint bundle'', which
    becomes a key ingredient of the conversion procedure for the
    non-scalar constraints. Unlike in the case of scalar second-class
    constraints, no Abelian conversion is possible in general.  Within
    the BRST framework, a systematic procedure is worked out for
    converting non-scalar second-class constraints into non-Abelian
    first-class ones. The BRST-extended system is quantized, yielding
    an explicitly covariant quantization of the original system. An
    important feature of second-class systems with non-scalar
    constraints is that the appropriately generalized Dirac bracket
    satisfies the Jacobi identity only on the constraint surface.  At
    the quantum level, this results in a weakly associative
    star-product on the phase space.
\end{minipage}
\end{center}

\thispagestyle{empty}
 \newpage

\begin{small}
  \tableofcontents
\end{small}
\section{Introduction}

The quantization problem is usually understood as that of constructing
a quantum theory for a given classical system, at the same time
preserving important properties of the system such as locality and global
symmetries. This additional requirement is crucial. Indeed, formally
one can always find a representation such that all the
constraints are solved, gauge symmetries are just shift symmetries, and
the Poisson bracket has the canonical form.  But in doing so one usually
destroys locality and global symmetries. It is the problem of
quantization of relativistic local field theories that initiated
the development of sophisticated quantization methods applicable to
systems with non-Abelian and open gauge algebras
\cite{Batalin:1977pb,deWit:1978cd,Fradkin:1978xi,Batalin:1981jr}.

From this point of view, the problem of quantizing curved phase space
appears as a problem of constructing quantization in a way that is
explicitly covariant with respect to arbitrary change of phase-space
coordinates. Given such a method (at the level of deformation
quantization at least) one can always find quantization in each
coordinate patch and then glue everything together. Similar to the
curved phase-space quantization problem is the one of quantizing
arbitrary constraint surface. Any surface can be represented by
independent equations (constraints) but in general only locally. In
fact, one can always assume that the surface is the zero locus of a
section of a vector bundle over the phase space. The quantization
problem for arbitrary constrained systems can then be reformulated as
the problem of constructing quantization that is explicitly covariant
with respect to the basis of constraints. In this paper, we restrict
ourselves to the case of second-class constraints and address the
problem of constructing a quantization scheme which is explicitly
covariant with respect to the change of phase-space coordinates and
constraint basis.

A general framework that allows to quantize second-class constraints
at the same footing as first-class ones is the well-known
\textit{conversion} -- the procedure that converts the original
second-class constraints into first-class ones by introducing extra
variables known as conversion variables. At least locally in the phase
space, any second-class constraints can be converted into Abelian
ones, and therefore the Abelian conversion is sufficient for most
applications. The situation changes drastically if one wants the
quantization to be explicitly covariant with respect to the change of
the constraint basis.  Indeed, by changing the constraint basis one
can always make the converted constraints non-Abelian. Additional price
one has to pay for covariance is the appearance of a connection in the
vector bundle associated with the constraints. This is reminiscent of the
quantization of systems with curved phase space, where phase-space
covariance requires introducing a symplectic connection on the phase
space. In fact, this is more than a coincidence.

The coordinate and constraint basis covariance appear to be intimately
related within the quantization methods developed
in~\cite{BFF,Fedosov-book} (see also~\cite{GL,BGL}). Indeed, the key
ingredient of these methods is the embedding of the system into the
cotangent bundle over its phase space. In the natural coordinate
system $x^i,p_i$, the embedding constraints $p_i=0$ are
non-scalar~\cite{GL}. In this example, the reparametrization
covariance in the original phase space translates into the covariance
with respect to the basis of constraints $p_i$.  In~\cite{BGL}, this
approach was extended to general second-class constrained systems with
constraints being scalar functions.

In this paper, we extend the method in~\cite{BGL} to the case where the
second-class constraint surface is an arbitrary symplectic
submanifold of the phase space, not necessarily defined by zero locus
of the set of any independent scalar functions. Considering the quantization
problem for the constrained systems whose classical dynamics evolves
on the constraint surface, one has to take care of the geometry of
the tubular neighborhood of the constrained submanifold. The geometry
of the entire phase space is irrelevant for this problem. In its turn, any tubular
neighborhood of the submanifold can be identified with the normal
bundle over the submanifold. For coisotropic submanifolds (first-class
constrained systems), the corresponding approach to quantization
was considered in~\cite{Cattaneo:2003dp}. It then follows that arbitrary constrained
submanifold can be considered as a zero locus of a section of the
appropriate vector bundle over the phase space. Moreover, in practical
physical problems, the second-class constraints can appear from the
outset as components of a section of some bundle over the original
phase space rather than scalar functions. This leads naturally to
the concept of constrained systems with non-scalar constraints.

By considering the original non-scalar constraints $\theta_\alpha$
at the same footing with the constraints $p_i$ determining the
embedding into $T^*\manM$, we achieve a globally defined description
for general second-class systems.  Using the appropriate
non-Abelian conversion procedure and subsequent BRST quantization,
we then arrive at the formulation of the quantum theory (at the
level of deformation quantization) that is explicitly covariant
under the reparametrizations of the original phase space
and under the changing the constraint basis. We note that the
non-scalar first-class constraints were also considered
in~\cite{Lyakhovich:2004kr} in a different framework.

The conventional approach to second-class systems is based on the
Dirac bracket -- a Poisson bracket on the entire phase space, which is
determined by constraints and for which the constraint surface is a
symplectic leaf. This allows considering the Poisson algebra of
observables as a Dirac bracket algebra of phase-space functions modulo
those vanishing on the constrained submanifold. From this point of
view, the quantization problem can be understood as that of quantizing
a degenerate Poisson bracket.  However, outside the constraint
surface, the Dirac bracket is not invariant under the change of the
constraint basis and therefore is not well-defined in the case of
non-scalar constraints.  The Dirac bracket bivector can be invariantly
continued from the constraint surface under certain natural
conditions, although the price is that the Jacobi identity is in
general satisfied only in the weak sense, i.e., on the constrained
submanifold.  In the non-Abelian conversion framework such a covariant
generalization of the Dirac bracket is naturally determined by the
Poisson bracket of observables of the converted system.  We note that
weak brackets were previously studied in various contexts
in~\cite{Batalin:1992,Batalin:2001fh,Deriglazov:1995ec,Lyakhovich:2004xd}.

At the quantum level, the lack of Jacobi identity for the covariant
Dirac bracket results in a phase-space star-product that is not
associative in general. Within the non-Abelian conversion approach developed in
the paper, this star-product naturally originates from the quantum
multiplication of BRST-invariant extensions of phase-space functions.
In the BRST cohomology, we obtain an associative star product which is
identified with the quantum deformation of the classical algebra of
observables (functions on the constraint surface). In particular, the
associativity of the phase-space star-product is violated only by the terms
vanishing on the constraint surface.

The quantization method developed in this paper can be viewed as an
extension of the Fedosov quantization
scheme~\cite{Fedosov:1994,Fedosov-book} to systems whose constrained
submanifolds are defined by non-scalar constraints and whose phase
spaces, as a result, carry a weak Poisson structure. We note that
gauge systems with a weak Poisson structure can be alternatively
quantized \cite{Lyakhovich:2004xd} using the Kontsevich formality
theorem.

\section{Geometry of constrained systems with locally defined constraints}
We consider a constrained system on a general symplectic manifold $(\omega ,
\manM )$.  The constrained system is defined on $\manM$ by
specifying a submanifold $\Sigma \subset \manM$ such that the
restriction $\omega|_\Sigma$ of the symplectic form to the
constraint surface has a constant rank. If $\Sigma$ is
coisotropic, the constrained system is called first-class. A
constrained system is called second-class if the restriction
$\omega|_\Sigma$ of the symplectic form is invertible on $\Sigma$.

Let us assume for a moment that $\Sigma$ is determined by
constraints $\theta_\alpha=0$ which are globally defined functions
on $\manM$, then $\pb{\theta_\alpha}{\theta_\beta}|_{\Sigma}=0$
($\pb{\theta_\alpha}{\theta_\beta}|_{\Sigma}$ is invertible) iff
the system is first- (respectively second-) class. The converse is
also true, but \textit{only locally}: if a constrained system is
first- (respectively second-) class then locally there exist
independent functions $\theta_\alpha$ determining constraint
surface $\Sigma$ by $\theta_\alpha=0$ and any such functions
satisfy $\pb{\theta_\alpha}{\theta_\beta}|_\Sigma=0$ (respectively
$\pb{\theta_\alpha}{\theta_\beta}|_\Sigma$ is invertible).

The dynamics of a constrained system on $\manM$ is assumed evolving
on the constraint surface $\Sigma\subset \manM$. At the quantum
level, a tubular neighborhood of $\Sigma$ gets involved in
describing dynamics. In its turn, it is a standard geometrical
fact that any such neighborhood is diffeomorphic to a vector
bundle over $\Sigma$. Indeed, in each neighborhood $U^{(i)}$ of a
point of $\Sigma$ one can pick a coordinate system
$x^a_{(i)},\theta^{(i)}_\alpha$ such that $\Sigma \cap U^{(i)}$ is
singled out by $\theta^{(i)}_\alpha=0$ and on the intersection of
two such neighborhoods $U^{(i)}$ and $U^{(j)}$
\begin{equation}
  x^a_{(i)}=X^a_{(ij)}(x_{(j)})\,,
\qquad \theta^{(i)}_\alpha=(\phi^{(ij)})^{\beta}_{\alpha}\theta^{(j)}_\beta
\end{equation}
with some functions $X^a_{(ij)}(x)$ and $\phi^{(ij)}(x)$.  Functions
$(\phi^{ij})^\alpha_\beta$ can be identified with transition functions
of a vector bundle $V^*(\Sigma)$ over $\Sigma$ (we use the notation
for a dual bundle to make notations convenient in what follows). Under
the identification of an open neighborhood of $\Sigma$ with the
vector bundle $V^*(\Sigma)$, coordinates $\theta_\alpha$ are
identified with constraint functions on $\manM$. In particular,
$\Sigma$ goes to the zero section of $V^*(\Sigma)$. Note that the
constraints $\theta_\alpha$, being understood as functions on $\manM$,
are defined only locally. If there exist globally defined constraints
then $V^*(\Sigma)$ is trivial.

It can be useful to pull back the vector bundle $V^*(\Sigma)$ to the
vector bundle $V^*(\manM)$ over $\manM$. Functions $\theta_\alpha$ are
then naturally identified with the components of a globally defined
section $\theta$ of $V^*(\manM)$. At the same time $\Sigma$
is nothing else than a submanifold of points where $\theta$
vanishes. These arguments motivate the following concept of a
constrained system:
\begin{definition}
  A constrained system with non-scalar constraints is a triple
  $(\manM,V^*(\manM),\theta)$ where $\manM$ -- symplectic manifold with
  a symplectic form $\omega$, $V^*(\manM)$ -- vector bundle over
  $\manM$, and $\theta$ is a fixed section of $V^*(\manM)$. It is
  assumed that vanishing points of $\theta$ are regular and form a
  submanifold $\Sigma \subset \manM$ (constraint surface) such that
  $\omega|_\Sigma$ has a constant rank.
\end{definition}
The definitions of first- and second-class constrained systems
still stand because they are formulated entirely in the
intrinsic terms of the constraint surface $\Sigma$, making use
only of the rank of $\omega|_\Sigma$ irrespectively to the way of
defining $\Sigma$.

Several comments are in order:

\begin{enumerate}
\item[(i)] Another possibility to consider arbitrary constraint
  surface keeping at the same time constraints globally defined
  functions is to use overcomplete sets of constraints (i.e.
  reducible constraints, in different terminology). However, depending
  on a particular system this can be a complicated task.  Moreover,
  even if the constraints are reducible it can also be useful to allow
  them to be non-scalar.
  
\item[(ii)] As we have seen, any submanifold $\Sigma \subset \manM$ can be
  represented as a surface of regular vanishing points of a section of
  a vector bundle over $\manM$. Note, however, that by taking
  arbitrary constraints $\theta^{(i)}_\alpha$ in each neighborhood
  $U^{(i)}$ one does not necessarily arrive at a vector bundle.
  Indeed, in the intersection $U^{(i)}\cap U^{(j)}$ one still has
\begin{equation}
\theta^{(i)}_\alpha=(\phi^{(ij)})^{\beta}_{\alpha}\theta^{(j)}_\beta
\end{equation}
But functions $(\phi^{(ij)})^\beta_\alpha$ are defined only up terms
of the form $(\chi^{(ij)})^{\beta\gamma}_\alpha\theta_\gamma$ with
$(\chi^{(ij)})^{\beta\gamma}_\alpha=-(\chi^{(ij)})^{\gamma\beta}_\alpha$.
As a consequence, functions $(\phi^{(ij)})^\beta_\alpha$ satisfy the
cocycle condition also up to terms proportional to $\theta$
\begin{equation}
  (\phi^{(ik)})_\alpha^\gamma (\phi^{(kj)})_\gamma^\beta=(\phi^{(ij)})_\alpha^\beta+\ldots\,.
\end{equation}
This means that only appropriately chosen constraints can be identified with components
of a section of a vector bundle over $\manM$. What differential geometry tells us is that
such a choice always exists.
\end{enumerate}

\section{Connections and symplectic structures on vector bundles}\label{subsec:symplectic}
In what follows we need some geometrical facts on the connections and symplectic structures
on the appropriately extended cotangent bundle over a symplectic
manifold. Let now $\manM$ be a symplectic manifold and $\cW(\manM) \to \manM$
be a symplectic vector bundle over $\manM$. Let also $e_A$ be a
local frame (locally defined basic sections of $\cW(\manM)$)
and $\mD$ be the symplectic form on the fibers of $\cW(\manM)$.
The components of $\mD$ with respect to $e_A$ are determined by $\mD_{AB}=\mD(e_A,e_B)$.

It is well known (see e.g.~\cite{Fedosov-book}) that any
symplectic vector bundle admits a symplectic connection.  Let
$\Gamma$ and $\nabla$ denote a symplectic connection and the
corresponding covariant differential in $\cW(\manM)$. The
compatibility condition reads as
\begin{equation}
\label{eq:compatible}
  \nabla \mD=0\,,  \qquad \d_i \mD_{AB} - \Gamma^C_{iA}\mD_{CB}-\Gamma^C_{iB}\mD_{AC}=0\,,
\end{equation}
where the coefficients $\Gamma^C_{iA}$ of $\Gamma$ are determined
as:
\begin{equation}
  \nabla e_A=dx^i \Gamma^C_{iA}e_C\,.
\end{equation}
It is useful to introduce the following connection 1-form:
\begin{equation}
\Gamma_{AB}=dx^i \Gamma_{AiB}\,, \qquad  \Gamma_{AiB}=\mD_{AC}\Gamma^C_{iB}\,.
\end{equation}
Then compatibility condition \eqref{eq:compatible} rewrites as
\begin{equation}
\label{eq:compatible-l}
d\mD_{AB}=\Gamma_{AB}-\Gamma_{BA}\,, \qquad  \d_i \mD_{AB} - \Gamma_{AiB}+\Gamma_{BiA}=0\,.
\end{equation}
As a consequence of the condition one arrives at the following
property of the connection 1-form $\Gamma_{AB}$:
\begin{equation}
\label{eq:dC-symm}
  d\Gamma_{AB}=d\Gamma_{BA}\,.
\end{equation}

Consider the following direct sum of vector bundles:
\begin{equation}
  \cE_0=\cW(\manM) \oplus T^*\manM\,,
\end{equation}
where $T^*\manM$ denotes a cotangent bundle over $\manM$.  Let
$x^i,p_j$ and $Y^A$ be standard local coordinates on $\cE_0$ ($x^i$ are
local coordinates on $\manM$, $p_j$ are standard coordinates on the
fibers of $T^*\manM$, and $Y^A$ are coordinates on the fibers of
$\cW(\manM)$ corresponding to the local frame $e_A$). Assume in
addition that $\manM$ is equipped with a closed 2-form $\omega$ (not
necessarily nondegenerate).

Considered as a manifold, $\cE_0$ can be equipped with the
following symplectic structure
\begin{multline}
    \label{eq:symplectic}
    \omega^{\,_{\cE_0}}=\pi^*\omega+2dp_i\wedge dx^i+\\
+\mD_{AB}dY^A\wedge
  dY^B+  d\Gamma_{AB} Y^AY^B -2 \Gamma_{AB}\wedge  dY^A Y^B\,,
\end{multline}
where $\pi^* \omega$ is the 2-form $\omega$ on $\manM$ pulled back
by the bundle projection $\pi\map\cE_0 \to \manM$.  One can directly
check that 2-form \eqref{eq:symplectic} is well defined.  Indeed,
it can be brought to the standard explicitly covariant form,
similar to that of the supersymplectic manifolds~\cite{Rothstein:1990bj}
\begin{equation}
  \omega^{\,_{\cE_0}}=\pi^*\omega+2dp_i\wedge dx^i
+ \mD_{AB}\nabla Y^A \wedge \nabla Y^B +  \mR_{AB} Y^A Y^B \, ,
\end{equation}
Here, $\nabla Y^A=  dY^A+\Gamma^A_CY^C \, ,$ and $\mR_{AB} =
\mR_{ij;\,AB} dx^i \wedge dx^j$ denotes the curvature of $\Gamma$:
\begin{multline}
  \mR_{ij;\,AB}=\mD_{AC}\mR^C_{ij\,B}~=\\
=~\mD_{AC}\left(\d_i \Gamma^C_{jB}-\d_j
\Gamma^C_{iB}+\Gamma^C_{iD}\Gamma^D_{jB}
-\Gamma^C_{jD}\Gamma^D_{iB}\right)~=\\
  =~\d_i \Gamma_{AjB}-\d_j
  \Gamma_{AiB}+\Gamma_{CiA}\mD^{CD}\Gamma_{DjB}-\Gamma_{CjA}\mD^{CD}\Gamma_{DiB}\,.
\end{multline}
The last equality follows from nondegeneracy of $\mD_{AB}$ and
compatibility condition~\eqref{eq:compatible-l}. Also, it is
straightforward to show that, the 2-form (\ref{eq:symplectic})
is exact, besides the first term:
\begin{equation}
\label{eq:primitive}
  \omega^{\,_{\cE}}~~=~~\pi^*\omega+d\left[2p_idx^i+
Y^A \mD_{AB} \nabla Y^B \right]
\end{equation}

Analyzing the structure in the r.h.s. of \eqref{eq:primitive} one
can see that an arbitrary (not necessarily symplectic) connection
$\st{0}{\Gamma}$ can be taken to construct the close $2$-form on
$\cE$ in~\eqref{eq:primitive}. It turns out that the resulting
$2$-form still has the structure~\eqref{eq:symplectic} with
$\Gamma$ given by
\begin{equation}
\label{eq:bare2symp}
  \Gamma_{AB}=\half(d\mD_{AB}+\st{0}{\Gamma}_{AB}+\st{0}{\Gamma}_{BA})\,.
\end{equation}
It is easy to see that connection $\Gamma$ is by construction
compatible with the symplectic structure $\mD$ for any connection
$\st{0}{\Gamma}$. In addition, if $\st{0}{\Gamma}$ was taken
symplectic it would bring $\Gamma=\st{0}{\Gamma}$.

The Poisson bracket on $\cE_0$ corresponding to the symplectic
form~\eqref{eq:symplectic} is determined by the following basic relations:
\begin{equation}
  \begin{aligned}
    \pb{p_i}{x^j}_{\cE_0}&=-\delta^j_i\,,\qquad
   & \pb{p_i}{p_j}_{\cE_0}&=\omega_{ij}(x)+\half \mR_{ij;AB}(x)Y^A\,Y^B\,,\\
    \pb{Y^A}{Y^B}_{\cE_0}&=\mD^{AB}(x)\,,\qquad
   & \pb{p_i}{Y^A}_{\cE_0}&= \Gamma^A_{iB}(x)Y^B\,,
  \end{aligned}
\end{equation}
with all the others vanishing:
$\pb{x^i}{Y^A}_{\cE_0}=\pb{x^i}{x^j}_{\cE_0}=0$.

\section{Embedding and conversion at the classical level}

\subsection{Embedding}
Consider a second-class constrained system $(\manM,V^*(\manM),\theta)$
with locally-defined constraints $\theta_\alpha$ (i.e. $\theta_\alpha$
are components of a section $\theta$ of $V^*(\manM)$ with respect to a
local frame $e^\alpha$). Let $T^*_\omega\manM$ be a cotangent bundle
equipped with the modified symplectic structure $2dp_i\wedge
dx^i+\pi_0^*\omega$, where $\omega$ is a symplectic form on $\manM$ and
$\pi_0 \map T^*_\omega\manM \to \manM$ is the canonical projection.

The embedding of $\manM$ into $T^*_\omega\manM$ as a zero section
is a symplectic map, i.e. a restriction of symplectic form $2dp_i\wedge
dx^i+\pi_0^*\omega$ to the submanifold $\manM$ is $\omega$.
Moreover, constrained system $(\manM,V^*(\manM),\theta)$
is equivalent to the constrained system
$(T^*_\omega\manM,W^*(\manM),\Theta)$, where $W^*(\manM)$ is a direct
sum $W^*(\manM)=T^*\manM \oplus V^*(\manM)$ considered as a vector
bundle over $T^*_\omega(\manM)$ and components of $\Theta$ with
respect to the local frame $dx^i,e^\alpha$ are $-p_i,\theta_\alpha$
(in other words, locally, the constraints are given by $-p_i=0$ and
$\theta_\alpha=0$).  Indeed, by solving constraints $-p_i=0$ one
arrives at the starting point constrained system. At this stage the
construction here repeats the one from~\cite{BGL} with the only
difference that constraints $\theta_\alpha$ are now defined only
locally.

\subsection{Non-Abelian conversion} Given second-class constraints
$\Theta_A$ one can always find an appropriate extension
of the phase space by introducing conversion variables $Y^A$ whose
Poisson bracket relations have the form $\pb{Y^A}{Y^B}=\mD^{AB}$
with $\mD^{AB}$ invertible. Then one can find converted constraints
$T_A$ in the extended phase space, satisfying
\begin{equation}
  \pb{T_A}{T_B}=U_{AB}^C T_C\,,\qquad  T_A\big|_{Y=0}=\Theta_A\,.
\end{equation}
The resulting first-class system with constraints $T_A$ is equivalent
to the original second-class one and is called converted system. For
second-class constraints that are scalar functions on the phase space
one can always assume the conversion to be Abelian, i.e. with
vanishing functions $U^C_{AB}$ (see~\cite{Batalin:1991jm} for a detailed
discussion of the conversion and the existence theorem for the Abelian
conversion).

For the non-scalar constraints one naturally wants to build a
converted constraints in the invariant way, i.e. independently of
a particular choice of the constraint basis. As we will see
momentarily this forces one to consider, in general, a non-Abelian
conversion.

To see this, one first needs to introduce conversion variables in
a geometrically covariant way. It is useful to take as conversion
variables the coordinates on the fibers of the bundle $W(\manM)$
dual to the bundle $W^*(\manM)=T^*\manM\oplus V^*(\manM)$
associated to constraints $\theta_\alpha,-p_i$. The phase space is
then
\begin{equation}
  \cE_0=T^*_\omega\manM\oplus W(\manM)\,,\qquad W(\manM)= V(\manM)\oplus T\manM
\end{equation}
We introduce unified notation $e_A$ and $Y^A$ for the local frame and
coordinates on the fibers of $W(\manM)$ respectively.  In the adapted
basis $Y^A$ split into $Y^i$ and $Y^\alpha$.

Given a connection $\bar\Gamma$ in $V(\manM)$ one can equip $W(\manM)$ with the following
fiberwise symplectic structure
\begin{equation}
  \mD_{ij}=\omega_{ij}\,, \qquad \mD_{i\alpha} =-\mD_{\alpha i}={\bar\nabla}_i\theta_{\alpha}=
  \d_i\theta_\alpha-{{\bar\Gamma}}_{i \alpha}^{\beta}\theta_\beta\,,\qquad
\mD_{\alpha\beta}=0\,.
\end{equation}
In what follows we also need the explicit form of its inverse $\mD^{AC},\, \mD^{AC}\mD_{CB}=\delta^A_B$
\begin{equation}
\label{covD}
  \begin{gathered}
\mD^{\alpha\beta}=\Delta^{\alpha\beta}\,,\qquad \mD^{i\beta}=-\omega^{il}\mD_{l\gamma}\Delta^{\gamma\beta}\,,\\
\mD^{ij}=\omega^{ij}-\omega^{ik}\mD_{k\alpha}\Delta^{\alpha\beta}\mD_{l\beta}\omega^{lj}\,,
  \end{gathered}
\end{equation}
where we introduced $\Delta^{\alpha\beta}$ as follows:
\begin{equation}
  \Delta^{\alpha\gamma}\Delta_{\gamma\beta}=\delta^\alpha_\beta\,,\qquad
\Delta_{\alpha\beta}=\mD_{i\alpha}\omega^{ij}\mD_{j\beta}\,.
\end{equation}
$\Delta$ is invertible on $\Sigma$ by assumption (recall that its
invertibility is a part of the defining property of second-class
constraints). It is then invertible in some neighborhood of
$\Sigma$ and we assume that it is invertible on the
entire~$\manM$. 

Note that $\mD^{ij}$ determines a bivector field on $\manM$ which
coincides on $\Sigma$ with the conventional Dirac bracket. The latter
bracket is not well-defined beyond $\Sigma$ if the constraints are not
scalars. The bracket determined by $\mD^{ij}$ in~\eqref{covD} can
therefore be understood as a covariant generalization of the Dirac
bracket to the case of non-scalar constraints. It is straightforward
to check that the covariant Dirac bracket satisfies Jacobi identity
modulo terms vanishing on $\Sigma$.

Furthermore, one can equip $W(\manM)$ with the symplectic connection
compatible with the fiberwise symplectic form. This is achieved as
follows.  First one picks a linear symplectic connection
${\bar\Gamma}_{\manM}$ on the symplectic manifold $\manM$ and equips
$W(\manM)$ with the direct sum connection $\st{0}{\Gamma}$ determined by
\begin{equation}
   \st{0}{\nabla}e_i=({\bar\Gamma}_{\manM})^j_i\, e_j\,,\qquad \st{0}{\nabla} e_\alpha
  =({\bar\Gamma})_\alpha^\beta \,e_\beta\,,
\end{equation}
where $\st{0}{\nabla}$ denotes the covariant differential determined by $\st{0}{\Gamma}$.
Given ``bare'' connection $\st{0}{\nabla}$ in $W(\manM)$ one then
arrives at the symplectic connection $\Gamma$
using~\eqref{eq:bare2symp}. In its turn the symplectic connection
in $W(\manM)$ determines a symplectic structure $\omega^{\cE_0}$
on $\cE_0$ in accordance to the general
formula~\eqref{eq:symplectic}. The associated
Poisson bracket reads as
\begin{equation}
  \begin{aligned}
    \pb{p_i}{x^j}&=-\delta_i^j\,,\qquad
   & \pb{p_i}{p_j}&= \omega_{ij}(x)+\half R_{ij;AB}(x)Y^A\,Y^B\,.\\
    \pb{Y^A}{Y^B}&=\mD^{AB}(x)\,,\qquad
    &\pb{p_i}{Y^A}&= \Gamma^A_{iB}(x)Y^B\,,
  \end{aligned}
  \label{eq:extended-PB}
\end{equation}
Here and in what follows we drop the superscript of the Poisson
bracket on the extended phase space whenever it can not lead to
confusions. Note that embedding of $T^*_\omega \manM$
into $\cE_0$ is symplectic. This implies that coordinates $Y^A$
can be treated as second-class constraints (they can also be understood as gauge
conditions for the converted system). Considered together with constraints
$\Theta_A$ they determine a constrained system on $\cE_0$ that is
equivalent to the original constrained system on $\manM$.

Since we are interested in the non-Abelian conversion, it is
preferable to work in terms of the BFV-BRST formalism from the very
beginning. To this end we introduce ghost variables $\cC^A$ and $\cP_A$
with the transformation law determined by that of components of a
section of $W(\manM)$ and $W^*(\manM)$ respectively. One can
consistently assume canonical Poisson bracket relations
\begin{equation}
  \pb{\cP_A}{\cC^B}=-\delta_A^B\,,
\end{equation}
and brackets between $\cC^A$ and $\cP_A$ and all other variables
vanishing. Note that in order for Poisson brackets between ghosts
and other variables to remain vanishing when passing from one
neighborhood to another momenta $p_i$ should transform
inhomogeneously. This means that the extended phase
space is $\cE=T^*_\omega(\Pi W(\manM))\oplus W(\manM)$ with
$W(\manM)$ in the second summand considered as a vector bundle
over $\Pi W(\manM)$. Here and below $\Pi$ indicates that the
Grassmann parity of the fibers of a vector bundle is reversed.
Note also that the extended phase space $\cE$ is not anymore a vector
bundle over $\manM$ because $p_i$ transform in an
inhomogeneous way.

In the BRST language the conversion problem can be formulated as
follows.  Given ``bare'' generating function $\bar\brst$ whose
expansion with respect to the ghosts variables starts with given
second-class constraints $\Theta_A$
\begin{equation}
\label{eq:conv-brst}
  \bar\brst=\cC^A \Theta_A+\ldots\,,\qquad \gh{\bar\brst}=1\,.
\end{equation}
The conversion implies finding BRST charge satisfying
\begin{equation}
  \label{eq:conversion}
\pb{\brst}{\brst}=0\,,\qquad\gh{\brst}=1\,,\qquad \brst|_{Y=0}=\bar\brst\,.
\end{equation}
Note that $\brst$ and $\bar\brst$ are assumed to be a globally defined
functions on the entire extended phase space and its submanifold determined by $Y^A=0$
respectively.

Now we describe conversion of the second-class constraints $\Theta_A=\{-p_i,\theta_\alpha\}$.
Taking into account their transformation properties a natural anzatz for a generating
function $\bar\brst$ is as follows
\begin{equation}
\label{eq:bar-brst}
  \bar\brst=-\cC^i p_i + \cC^\alpha\theta_\alpha+
\cC^i({\bar\Gamma})_{i\beta}^\alpha \cC^\beta \cP_\alpha\,.
\end{equation}
Indeed, the nonlinear in ghosts term coming from the
transformation law for $p_i$ is compensated by the term coming
from inhomogeneous contribution in the transformation law for the
connection coefficients.  This is exactly the point. In order for
the generating function $\bar\brst$ as well as BRST charge $\brst$
to be globally defined functions, one needs to introduce the terms
nonlinear in ghosts. In terms of constraints, this implies that
the conversion is non-Abelian.

\subsection{Existence and construction of the classical BRST charge}
In the standard BFV-BRST formalism the BRST charge and BRST invariant
observables are constructed by expanding in homogeneity degree in
ghost momenta.  The existence of a nilpotent BRST charge is ensured by
Homological Perturbation Theory~\cite{Henneaux:1985kr}
with the relevant operator being Koszul-Tate differential
associated with the constraints.  At the same time, within the
Abelian conversion procedure the effective first-class constraints,
BRST charge, and BRST-invariant observables are constructed by
expanding in homogeneity degree in conversion variables and all these
quantities are to be found order by order in these variables.

In the case of non-Abelian conversion it is then natural
to take as an expansion degree the total homogeneity in ghost
momenta $\cP_A$ and conversion variables $Y^A$:
\begin{equation}
\label{eq:deg-class}
  \deg{Y^A}=\deg{\cP_A}=1\,,\qquad \deg{x^i}=\deg{p_i}=\deg{\cC^A}=0\,.
\end{equation}
Accordingly, $\brst$ decomposes as
\begin{equation}
\label{eq:bc1.5}
  \brst=\sum_{s=0} \brst_s\,, \qquad \brst_0=\cC^\alpha \theta_\alpha-\cC^ip_i\,,\quad
\brst_1=\cC^i({\bar\Gamma})_{i\beta}^\alpha \cC^\beta \cP_\alpha+\ldots\,,
\end{equation}
where we have explicitly kept the term from the first order
contribution which is needed for covariance. The required BRST charge
satisfying~\eqref{eq:conversion} is to be constructed order by order
in the degree. To this end one first needs to satisfy the master
equation to zeroth order in the degree which implies finding
$\brst_1$. A ``minimal'' form of $\brst_1$ which satisfies master
equation to the zeroth order can be taken as
\begin{equation}
\label{eq:bc1}
  \brst_1=\cC^i({\bar\Gamma})_{i\beta}^\alpha \cC^\beta \cP_\alpha-
\cC^A\mD_{AB}Y^B\,.
\end{equation}

In constructing BRST charge it is also useful to restrict ourselves to
the following class of phase-space functions: let $\aA^0$ be the space
of formal power series in $Y^A$, ghosts $\cC^A$, and ghost momenta
$\cP_\alpha$ with coefficients being smooth functions in $x^i$. In
other words we forbid dependence on $p_i$ and $\cP_i$.  The space
$\aA^0$ is closed under the multiplication and the Poisson bracket
(both operations can be naturally defined for formal power series).
Algebra $\aA^0$ decomposes with respect to the
degree~\eqref{eq:deg-class} as $\aA^0=\oplus_{s\geq 0}\aA^0_s$ so that an
element of $\aA^0_s$ has the form
\begin{equation}
a=\sum_{p\geq 0, q\geq 0}^{p+q=s} (a_{pq})_{A_1\ldots A_p}^{\alpha_1\ldots\alpha_q}
Y^{A_1}\ldots Y^{A_p}\,\cP_{\alpha_1}\ldots\cP_{\alpha_q}\,,\qquad a_{pq}=a_{pq}(x,\cC)\,.
\end{equation}

Since the BRST charge and BRST invariant observables are to be
constructed by expanding in the degree~\eqref{eq:deg-class} the
lowest degree term $-\delta$ in the expansion of $\pb{\brst}{\cdot}$
plays a role of the nilpotent operator determining homological perturbation theory.
Considered acting on elements from
$\aA^0$, operator $\delta$ is completely determined by $\bar\brst$ and is given by
degree $-1$ operator
\begin{equation}
  \delta=\cC^A\dl{Y^A}+\theta_\alpha\dl{\cP_\alpha}\,.
\end{equation}
It is therefore a sum of standard Koszul--Tate operator
$\delta_K=\theta_\alpha\dl{\cP_\alpha}$ associated with original
constraints $\theta_\alpha$ and the operator $\cC^A\dl{Y^A}$ which
determines a homological perturbation theory in the Abelian
conversion framework and in the Fedosov quantization.

To proceed with the conversion we need to introduce a version of
the contracting homotopy operator determined by
\begin{equation}
  \delta^* f_{pq}=\frac{1}{p+q}Y^A\dl{\cC^A} f_{pq} \,,\quad p+q\neq 0\,,\quad\qquad \delta^*
  f_{00}=0\, , \qquad \delta^{* \, 2} \equiv 0
\end{equation}
for an element $f_{pq}\in\aA^0$ which is homogeneous in $\cC^A$ and $Y^A$ of orders $p$ and $q$
respectively. Operators $\delta$ and $\delta^*$ satisfy
\begin{equation}
\label{eq:ch}
  \delta^*\delta a+\delta\delta^* a=a-a|_{\cC=Y=0}\,.
\end{equation}
\begin{prop}\label{prop:exist-class}
  There exists a classical BRST charge $\brst$, $\gh{\brst}=1$ satisfying master equation
  $\pb{\brst}{\brst}=0$, boundary conditions \eqref{eq:bc1.5},\eqref{eq:bc1}, and such
  that $\brst_s\in \aA^0_s$ for $s\geq 2$.  In addition, given $\brst_0$
  and $\brst_1$ such a BRST charge is unique provided
  $\delta^*\brst_s=0$ for all $s\geq 2$.
\end{prop}
\begin{proof}
The Poisson bracket on $\cE$ can also be expanded with respect to
the degree as
\begin{equation}
\pb{\cdot}{\cdot}=\pb{\cdot}{\cdot}_{-2}+\pb{\cdot}{\cdot}_{-1}+\pb{\cdot}{\cdot}_0
+\pb{\cdot}{\cdot}_2
\end{equation}
(terms with other degrees vanish) where each term is a bilinear
first order differential operator of definite degree. In particular
\begin{equation}
\pb{f}{g}_{-2}=f\dr{Y^A}\mD^{AB}\dl{Y^B}g\,.
\end{equation}

The master equation at order $n$ in degree implies,
\begin{equation}
\label{eq:n-th}
 \pb{\brst_0}{\brst_{n+2}}_{-2}+\pb{\brst_0}{\brst_{n+1}}_{-1}+\pb{\brst_1}{\brst_{n+1}}_{-2}+B_n=0
\end{equation}
where $B_n$ depends on $\brst_s$ with $s\leq n$ only and is given explicitly by
\begin{equation}
  B_n=\sum_{0\geq p,q \geq n}^{p+q+s=n}\pb{\brst_p}{\brst_q}_s\,.
\end{equation}
In fact, the first term in \eqref{eq:n-th} vanishes because $\brst_0$ doesn't depend on $Y$
and the equation takes the form
\begin{equation}
\label{eq:n-th-short}
  \delta \brst_{n+1}=B_n\,.
\end{equation}
This equation can always be solved by
$\brst_{n+1}=\delta^* B_n$ using~\eqref{eq:ch}, $B_n|_{\cC=Y=0}=0$, and the consistency condition $\delta
B_n=0$. The later is fulfilled provided the master equation holds to lowest
orders, i.e. that $\brst_s$ for $s\leq n$ are such that
\begin{equation}
\{{\st{(n)}{\brst}},{\st{(n)}{\brst}}\}\in \bigoplus_{s\geq n}\aA^0_{s}\,,
\qquad \st{(n)}{\brst}=\sum_{s=0}^n\brst_s\,.
\end{equation}
Indeed, consider the following identity
\begin{equation}
  \pb{\st{(n)}{\brst}}{\pb{\st{(n)}{\brst}}{\st{(n)}{\brst}}}=0\,.
\end{equation}
Next, observe that $\pb{\st{(n)}{\brst}}{\st{(n)}{\brst}}=B_n+\ldots$
with dots denoting terms from $\aA^0_{\geq n+1}$ and, finally, check that to order
$n-1$ in the degree this identity gives $\delta B_n=0$.

This solution for $\brst_{n+1}$ obviously belongs to $\aA^0$ and satisfies $\delta^*\brst_{n+1}=0$. Conversely,
equation \eqref{eq:n-th-short} has a unique solution $\brst_{n+1}$ satisfying
$\brst_{n+1}\in\aA^0_{n+1}$, $\delta^*\brst_{n+1}=0$,
and $\gh{\brst_{n+1}}=1$.
\end{proof}

\subsection{Classical observables and weak Dirac bracket}\label{sec:observ-class}
We show that observables of the original system on $\manM$ are
isomorphic to observables of the BFV-BRST system on $\cE$. The
latter are understood as cohomology of the adjoint action
\begin{equation}
  Q=\pb{\brst}{\cdot}
\end{equation}
of the BRST charge.
\begin{prop}\label{prop:lift-class}
Let $f_0=f_0(x,\cC)$ be any $Y$ and $\cP$-independent function.
Then there exists $f\in\aA^0$ such that
\begin{equation}\label{eq:lift-cond}
  \pb{\brst}{f}=0\,,\quad f|_{Y=\cP=0}=f_0\,, \quad \gh{f}=\gh{f_0}\,.
\end{equation}
If in addition $\delta^*(f-f_0)=0$ and $\gh{f}\geq 0$ then $f$
is a unique BRST invariant extension of $f_0$. Moreover, if $f,\tilde f\in\aA^0$  both
satisfy~\eqref{eq:lift-cond} with the same function $f_0$, then
$f-\tilde f=Qh$ for some function $h\in\aA^0$.
\end{prop}
\begin{proof}The proof is standard and follows by expanding $Qf=0$ with respect to
  degree~\eqref{eq:deg-class} and using the fact that
  $\delta$-cohomology is trivial in nonzero degree. The later
  statement obviously holds provided cohomology of the
  standard Koszul--Tate operator $\delta_K=\theta_\alpha\dl{\cP_\alpha}$
  vanishes in nonzero degree in $\cP_\alpha$. Locally, operator $\delta_K$ is
  known to have vanishing cohomology in nonzero degree provided
  constraints $\theta_\alpha$ satisfy standard regularity assumptions.
  This also holds globally as can be shown by using suitable partition of unity.
\end{proof}

Let $f_0$ and $g_0$ be two inequivalent observables of the
original system, i.e.  $f_0|_\Sigma-g_0|_\Sigma\neq 0$. It then
follows from the explicit form of $\brst$ that their BRST
invariant extensions $f$ and $g$ determined by
Proposition~\bref{prop:lift-class} are not equivalent, i.e.,
$f-g\neq\pb{\brst}{h}$ for any $h$. This means that
observables of the original system  are observables of the
BFV-BRST system. In fact, one can show that these systems are
equivalent in the sense that the Poisson algebra of inequivalent
observables of the original system (i.e. the algebra of functions
on $\Sigma$ equipped with the Poisson bracket) is isomorphic to
the Poisson algebra of ghost number zero BRST cohomology of the
BFV-BRST system. Now we restrict ourselves to a little bit weaker
equivalence statement. Namely we show that this holds for BRST
cohomology evaluated in $\aA^0$ ($Q$ obviously maps $\aA^0$ to
itself).
\begin{prop}
  Let $f$ be an arbitrary function from $\aA^0$ satisfying $Qf=0$. Then
  $f=Qh$ for some $h$ iff
  $f\big|_{\Sigma}=f\big|_{\theta_\alpha=\cC^A=Y^A=\cP_\alpha=0}=0$.
\end{prop}
\begin{proof}
Let $f_0=f|_{Y=\cP=0}$. Condition $f|_{\Sigma}=f_0|_\Sigma=0$ implies that
there exist $f_0^\alpha(x)$ and $f_{0\,A}(x,\cC)$ such that
\begin{equation}
  f_0=\theta_\alpha f_0^\alpha
 +\cC^A f_{0A}
\end{equation}
and their transformation properties can be assumed to be those of
sections of $V(\manM)$ and $W^*(\manM)$ respectively.
One can then check that
\begin{equation}
(Qh)|_{Y=\cP=0}=f_0\,,\qquad h=-\cP_\alpha f_0^\alpha-Y^Af_{0A}
\end{equation}
because $f_0=-\delta h$ and $(Q h)|_{\aA^0_0}=-\delta h$ for $h\in\aA^0_1$.  Proposition
\bref{prop:lift-class} then implies that there exists $h^\prime\in\aA^0$ such
that $f=Q(h+h^\prime)$.
\end{proof}
To summaries we have
\begin{theorem}\label{thm:coh-class}
The BRST cohomology of $Q=\pb{\brst}{\cdot}$ evaluated in $\aA^0$
are given by
\begin{equation}
  \begin{aligned}
    H^n(Q,\aA^0)&=C^\infty(\Sigma)\qquad &n&=0\,,\\
    H^n(Q,\aA^0)&=0\qquad \qquad &n&\neq 0\,.
\end{aligned}
\end{equation}
\end{theorem}

The fact that all the physical observables can be taken elements
of $\aA^0$ suggests to consider $\aA^0$ as a fundamental object
replacing algebra of functions on the entire extended phase space.
This can be consistently done in spite of the fact that the BRST
charge $\brst$ and the ghost charge $\cG=\cC^A\cP_A$ do not belong
to $\aA^0$.  Indeed, from a more general point of view, a
classical BFV-BRST system is determined by (i) Poisson algebra
with not necessarily nondegenerate Poisson structure, which is
also graded with the ghost degree (ii) Odd nilpotent BRST
differential $Q$ of ghost number $1$ that differentiates both the
product of functions and the Poisson bracket and (iii)
differential $V$ (determining evolution) of zero ghost number
which differentiates both the product and the Poisson bracket and
satisfies $\commut{Q}{V}=0$. The standard Hamiltonian BFV-BRST
system fits this definition with $Q=\pb{\brst}{\cdot}$ and
$V=\pb{H}{\cdot}$ with $H$ denoting Hamiltonian. Such a
generalization of the Hamiltonian BFV-BRST theory was recently
studied in~\cite{Lyakhovich:2004xd}. Note also that in the
Lagrangian context this corresponds to theories described by
BRST differential not necessarily generated by a master action
and an antibracket. Theories of this type were recently considered
in~\cite{BGST}.

From this slightly more general point of view, the Poisson algebra
$\aA^0$ is a BFV-BRST system because $Q$ and ghost number operator
preserve $\aA^0$. The notion of generalized BFV-BRST
system can be extended to the quantum case by replacing the
Poisson algebra with the star-product algebra. It can also be
generalized further in the sense that the bracket can be allowed
to satisfy Jacobi identity only up to $Q$-exact terms as well as
$V$ can preserve the bracket only
weakly~\cite{Lyakhovich:2004xd}.

Let us give some further comments concerning the Poisson bracket of BRST
observables.  In the case where $\brst$ is Abelian (see~\cite{BGL} for
detailed discussion of this case) Proposition~\bref{prop:lift-class}
establishes an isomorphism between the algebra of functions of $x^i$
and functions of $x^i,Y^A$ satisfying $\pb{\brst}{\cdot}=0$ and
$\delta^*\cdot=0$. The later algebra (understood as a subalgebra in
$\aA^0)$ is closed under the Poisson bracket in $\aA^0$. The Poisson
bracket in this algebra determines a Poisson bracket on $\manM$ that
can be easily seen to coincide with the Dirac bracket associated to
second-class constraints $\theta_\alpha$.

In the present case $\brst$ explicitly depends on
$\cP_\alpha$ and one is forced to consider $\delta^*$
and $\pb{\brst}{\cdot}$-closed functions from $\aA^0$ which are
now allowed to depend also on $\cC^A$ and $\cP_\alpha$.  However,
this algebra is not anymore closed under the Poisson bracket and
therefore a direct counterpart of the Dirac bracket fails to
satisfy Jacobi identity outside $\Sigma$ in this case. Indeed,
finding unique lifts $f,g\in\aA^0$ of two phase-space functions
$f_0$ and $g_0$, evaluating their Poisson bracket, and putting
$Y=\cP=0$ one finds a bracket on $\manM$ which coincides with the
standard Dirac bracket when $\theta_\alpha=0$.
Explicitly, the bracket reads
\begin{equation}
  \pb{f_0}{g_0}_D=\d_i{f_0}\mD^{ij}\d_j{g_0}=
\d_i{f_0}\,\omega^{ij}\d_j g_0-\d_i f_0 \,\omega^{il}\bar\nabla_l\theta_\alpha
\Delta^{\alpha\beta}\bar\nabla_k\theta_\beta\omega^{kj}\d_j g_0\,,
\label{CDB}
\end{equation}
where
\begin{equation}
\bar\nabla_i\theta_\alpha=\d_i\theta_\alpha-\bar\Gamma_{i\alpha}^\beta\theta_\beta\,,\qquad
\Delta^{\alpha\gamma}\Delta_{\gamma\beta}=\delta^\alpha_\beta\,,\quad
\Delta_{\alpha\beta}=\bar\nabla_i\theta_\alpha\omega^{ij}\bar\nabla_j\theta_\beta\,.
\end{equation}
This bracket can be considered as a direct generalization of the
standard Dirac bracket. Unlike the later this generalized bracket does
not depend on the choice of constraint basis and therefore is
well-defined outside $\Sigma$ in the case of non-scalar constraints.
The Jacobi identity for the bracket~\eqref{CDB} is violated
by the terms proportional to the curvature $\bar R^\alpha_{ij\,\beta}$
of the connection $\bar\nabla$ and to the constraints $\theta_\alpha$.
So it is inevitably a weak bracket if the bundle $V({\manM})$ does not
admit flat connection.

\subsection{Dirac connection}
As we have seen the construction imposes no constraints
on the connection $\Gamma^A_{iB}$ entering the Poisson bracket
on $\cE$ but the compatibility with the symplectic form $\mD_{AB}$.
Symplectic connection always exists and can be obtained starting from
arbitrary connection in $W(\manM)$, e.g., using~\eqref{eq:bare2symp}
Let us, nebertheless, give an explicit form of the particular
symplectic connection which as we are going to see also has
some additional properties.

To this end let us consider an explicit form of the compatibility condition $\nabla\mD=0$
\begin{equation}
\label{eq:symp-cond}
\begin{gathered}
\d_i\omega_{jk}-\Gamma_{jik}+\Gamma_{kij}=0\,,\\
\d_i{\bar\nabla}_j\theta_\alpha-\Gamma_{ji\alpha}+\Gamma_{\alpha ij}=0\,,\\
\Gamma_{\alpha i \beta}-\Gamma_{\beta i \alpha}=0\,.
\end{gathered}
\end{equation}
The solution which is compatible with the transformation properties
and which is in some sense a minimal choice reads as
\begin{equation}
\label{eq:D0}
  \begin{aligned}
    \Gamma_{\alpha i\beta}&=0\,,\qquad
    &\Gamma_{ji\alpha}&=\mD_{j\beta}({\bar\Gamma})^\beta_{i\alpha}\,,
\\
    \Gamma_{ijk}&=({\bar\Gamma}_\manM)_{ijk}\,,\qquad
    &\Gamma_{\alpha ij}&=-\bar\nabla_i \bar\nabla_j\theta_\alpha\,,
\end{aligned}
\end{equation}
where $\bar\nabla_i\bar\nabla_j\theta_\alpha=\bar\nabla_i\mD_{j\alpha}=\d_i\mD_{j\alpha}
-\Gamma_{i\alpha}^\beta\mD_{j\beta}$.

It is easy to see that if $V(\manM)$ is trivial and one takes
$\bar\Gamma=0$ then \eqref{eq:D0} coincides with the Dirac connection
introduced in~\cite{BGL}.  In fact, connection
\eqref{eq:D0} possesses similar properties with respect to
the weak Dirac bracket. To see this let us write down this
connection in terms of the coefficients with upper indices
\begin{equation}
  \begin{aligned}
    \Gamma^j_{ik}&=\omega^{jl}\left(({\bar\Gamma_\manM})_{lik}+
\mD_{l\gamma}\Delta^{\gamma\alpha}\hat\nabla_i\hat\nabla_k\theta_\alpha\right)\,,
\qquad
& \Gamma^j_{i\alpha}&=0\,,
\\
\Gamma^\alpha_{ij}&=-\Delta^{\alpha\beta}\hat\nabla_i\hat\nabla_j\theta_\beta\,,
\qquad
&\Gamma_{i\alpha}^\beta&=\bar\Gamma_{i\alpha}^\beta\,,
\end{aligned}
\end{equation}
where $\hat\nabla_i\theta_\alpha=\bar\nabla_i\theta_\alpha$ and
$\hat\nabla_i\hat\nabla_j\theta_\alpha=
\d_i\mD_{j\alpha}-\bar\Gamma_{i\alpha}^\beta\mD_{j\beta}-(\bar\Gamma_\manM)_{ji}^k\mD_{k\alpha}$.
Connection $\Gamma$ in $W(\manM)$ determines a connection
$\Gamma_\mD$ in $T\manM$ whose coefficients are $\Gamma^j_{ik}$.
It follows from $\nabla\mD^{AB}=0$ and $\Gamma^j_{i\alpha}=0$ that
\begin{equation}
  (\nabla_\mD)_i\mD^{jk}=\d_i\mD^{jk}+\Gamma_{il}^j\mD^{lk}+\Gamma_{il}^k\mD^{jl}=0\,,
\end{equation}
which means that the Dirac bivector is covariantly constant with
respect to the connection $\Gamma_\mD$. One then concludes
that $\Gamma_\mD$ can be considered as a generalization of the Dirac
connection introduced in~\cite{BGL}.

 Note that $\Gamma_\mD$ is in
general not symmetric and its torsion is proportional to the curvature
of $\bar\Gamma$. On the constraint surface this connection coincides
with the Dirac connection in~\cite{BGL}. Similar arguments then show
that $\Gamma_\manM$ can be restricted to $\Sigma$ and its restriction
is a symplectic connection on $\Sigma$ considered as a symplectic
manifold.

In the construction of $\Gamma$ there is an ambiguity described by
an arbitrary $1$-form with values in symmetric tensor
product of the bundle $W^*(\manM)$ (i.e. arbitrary connection has the
form $\Gamma_{AiB}=\Gamma^{\mathrm{fixed}}_{AiB}+\gamma_{AiB}$,
with $\gamma_{AiB}-\gamma_{BiA}=0$). One can try to find  
additional conditions in order to fix the ambiguity in the connection.
In particular, to find an invariant criterion which allows to separate
connections compatible with the Dirac bracket. 

It turns out that it is possible to formulate a condition of this type
by analyzing the conversion procedure. To see this we note that the
term in $\brst_2$ of the form $\cC^i\gamma_{AiB}Y^AY^B$ can be
absorbed into the redefinition of $p_i$ which in turn leads to the
adjustment of the symplectic connection
$\Gamma_{AiB}\to\Gamma_{AiB}+\gamma_{AiB}$.  It is then natural to
choose the connection such that the respective contribution to
$\brst_2$ vanishes, i.e. the connection which is not modified by the
conversion. For $\brst$ satisfying conditions of the second part of
Proposition~\bref{prop:exist-class} this implies that
\begin{equation}
  \left(\delta^*\pb{-\cC^ip_i+
\cC^i\bar\Gamma_{i\alpha}^\beta \cC^\alpha\cP_\beta}{C^A \mD_{AB}Y^B}\right)\Big|_{\cC^\alpha=0}=0\,.
\end{equation}
This gives the following conditions on $\Gamma$
\begin{equation}
\label{eq:add-cond}
  \begin{gathered}
\d_i\omega_{jk}-\Gamma_{jik}+\d_k\omega_{ji}-\Gamma_{jki}+\Gamma_{ijk}+\Gamma_{kji}=0\,,\\
\d_i\mD_{j\alpha}-\mD_{i\beta}\bar\Gamma_{j\alpha}^\beta
-\Gamma_{ji\alpha}+\Gamma_{ij\alpha}+\Gamma_{\alpha ji}=0\,,\\
\Gamma_{\alpha i\beta}+\Gamma_{ \beta i\alpha}=0\,,
\end{gathered}
\end{equation}
If one takes $\Gamma_{ijk}=(\bar\Gamma_\manM)_{ijk}$ where $(\bar\Gamma_\manM)_{ijk}$ are coefficients
of a fixed symmetric symplectic connection on $\manM$ then 
equations~\eqref{eq:symp-cond} and \eqref{eq:add-cond} have a unique
solution with $\Gamma_{ijk}=({\bar\Gamma_\manM})_{ijk}$. It is given explicitly by
\begin{equation}
  \begin{aligned}
    \Gamma_{\alpha i\beta}&=0\,,\qquad
    &\Gamma_{ji\alpha}&=\mD_{j\beta}({\bar\Gamma})^\beta_{i\alpha}
+\frac{1}{3}{\bar R}_{ji\alpha}^\beta\theta_\beta\,,\\
    \Gamma_{ijk}&=({\bar\Gamma}_\manM)_{ijk}\,,\qquad
    &\Gamma_{\alpha ij}&
=-\bar\nabla_i \bar\nabla_j\theta_\alpha-
\frac{1}{3}{\bar R}_{ij\alpha}^\beta\theta_\beta\,,
\end{aligned}
\end{equation}
Note that consistency of~\eqref{eq:symp-cond} and \eqref{eq:add-cond} together with $d\omega=0$
requires $\Gamma_{ijk}-\Gamma_{ikj}=0$.

This connection differs form the one in~\eqref{eq:D0} by the terms proportional
to ${\bar R}_{ij\alpha}^\beta\theta_\beta$. It also determines a connection
$\Gamma^\prime_\mD$ on $\manM$ which coincides with $\Gamma_\mD$ on $\Sigma$.
In particular, $\Gamma^\prime_\mD$ is compatible with the Dirac bracket
only weakly in general.

\section{Conversion at the quantum level}
\subsection{Quantization of the extended phase space}\label{sec:quant}
At the quantum level we concentrate on the algebra
$\qA^0=\aA^0\tensor [[\hbar]]$ and its extension $\qA$ obtained by allowing dependence on $p_i$ and
$\cP_i$ through the following combinations (see~\cite{BGL} for details)
\begin{equation}
  \P=\cC^i(-p_i+\bar\Gamma^\alpha_{i\beta}\,\cC^\beta\cP_\alpha)\,,\qquad \G=\cC^i\cP_i\,.
\end{equation}
A general element of $\qA$ has the form
\begin{equation}
  a=\P^r\G^sa_0\,, r=0,1,\quad s=0,1,\ldots ,\dim(\manM)\,,\quad a_0\in\qA^0
\end{equation}
The algebra $\qA$ is closed under ordinary multiplication and
the Poisson bracket. Moreover, it can be directly quantized.
To this end one first quantize $\qA^0$ by introducing Weyl star-product
according to
\begin{multline}
\label{eq:Weyl-star-product}
(a\star b)(x,Y,\cC,\cP,\hbar)~~=\\
=~~\{(a(x,Y_1,\cC_1,\cP_2,\hbar) exp ( -\frac{i\hbar}{2} (\mD^{AB}
\dr{Y_1^A} \dl{Y_2^B}-\dr{\cC_1^\alpha}\dl{\cP^2_\alpha}-
\dr{\cP^1_\alpha}\dl{\cC_2^\alpha}))\\
b(x,Y_2,\cC_2,\cP_2,\hbar)\}
{\bigr|}_{Y_1=Y_2=Y,\,\cC_1=\cC_2=\cC,\,\cP_1=\cP_2=\cP} \,\,,
\end{multline}
where $\cP$ stands for $\cP_\alpha$ only. This star product can also be extended
from $\qA^0$ to $\qA$. Here we give only those formulas which we really need in what
follows (we refer to~\cite{BGL,GL} for further details of such an extensions):
\begin{equation}
  \begin{gathered}
    \frac{i}{\hbar}\commut{\P}{a}=\cC^i(\dl{x^i}-\bar\Gamma_{i\alpha}^\beta\cC^\alpha\dl{\cC^\beta}
    +\bar\Gamma_{i\beta}^\alpha\cP_\beta\dl{\cP_\beta} -\Gamma_{iA}^B
    Y^B\dl{Y^A})a\,,\quad a\in \qA^0\,,\\
    \frac{i}{\hbar}\commut{\P}{\P}=-i\hbar\cC^i\cC^j(\omega_{ij}+\bar
    R_{ij\alpha}^\beta\cC^\alpha\cP_\beta+\half \mR_{ij\,AB}
    Y^AY^B)\,,\\
\frac{i}{\hbar}\commut{\G}{a}=-\cC^i\dl{\cC^i}+\cP_i\dl{\cP_i}\,,\quad a\in\qA^0\,.
\end{gathered}
\end{equation}
Note that these relations are enough to consistently consider
$\qA^0$ as a star-product algebra underlying the BFV-BRST system
at the quantum level in the sense described
in~\bref{sec:observ-class}.

On $\qA$ we introduce the following degree
\begin{equation}
  \deg{Y^A}=\deg{\cP_A}=\deg{p_i}=1\,,\qquad \deg{\cC^A}=\deg{x^i}=0\,,
\qquad \deg{\hbar}=2\,.
\end{equation}
One then decomposes $\qA^0$ and $\qA$ with respect to the degree as
\begin{equation}
  \qA^0=\bigoplus_{s=0}\qA^0_s
\end{equation}
and similarly for $\qA$. The star product also decomposes into homogeneous
components with respect to degree
\begin{equation}
  \star=\star_0+\star_1+\star_2+\ldots
\end{equation}
In particular, $\star_0$ contains ordinary product, Weyl product in
the sector of $Y$ variables, and the component of the product which
takes $\P$ with itself into $-\imath\hbar\cC^i\cC^j\omega_{ij}$.

Let us note that the choice of the degree is not unique. The one
we are using is convenient for general proofs but perhaps is not
the most suitable for computations because it is not preserved by
the star product in $\hat\aA^0$. From this point of view one can
consider another degree for which $\deg \cC^A=1$ and gradings of
other variables left unchanged.

\subsection{Quantum BRST charge}
Now we are going to show the existence of the quantum BRST charge
satisfying
\begin{equation}
\label{eq:q-master}
  \commut{\hat\brst}{\hat\brst}=0\,,\qquad \gh{\hat\brst}=1\,,
\end{equation}
together with the condition $\hat\brst|_{\hbar=0}=\brst$.
Here and below $\commut{\cdot}{\cdot}$ stands for the graded
commutator with respect to the star-multiplication in
$\hat\aA$, which is also decomposed into homogeneous component
with respect to the degree. A degree $s$ component of the
commutator is denoted by $\commut{\cdot}{\cdot}_s$.

It follows from the standard deformation theory and the vanishing of
$Q$-cohomology in nonzero ghost number that quantum BRST
charge exists. However, instead of deforming the classical BRST charge
we construct the quantum one from scratch.  To this end we show that
the quantum master equation \eqref{eq:q-master} has a solution
satisfying the following boundary conditions
\begin{equation}
\label{eq:q-boundary}
  \hat\brst_0=\cC^\alpha\theta_\alpha\,,\qquad \hat\brst_1=\P-\cC^A \mD_{AB}Y^B=
-\cC^ip_i+\cC^i\bar\Gamma^\alpha_{i\beta}\cC^\beta\cP_\alpha-\cC^A \mD_{AB}Y^B\,.
\end{equation}
\begin{prop}
  Equation~\eqref{eq:q-master} has a solution satisfying boundary
  condition~\eqref{eq:q-boundary} and $\hat\brst_s\subset\qA^0_s$
  for $s\geq 2$. Under the additional condition
  $\delta^*\hat\brst_s=0,\,\,s\geq 2$ the solution is unique.
\end{prop}
\begin{proof}
  Prof is completely standard once degree is prescribed. The only
  thing to check is that with boundary
  conditions~\eqref{eq:q-boundary}, the master equation holds at orders
  $0,1$ and $2$ which is straightforward. The rest follows by induction
  using that
\begin{equation}
\commut{\hat\brst_0}{a}_0=0\,,\qquad
\frac{1}{i\hbar}\left(\commut{\hat\brst_0}{a}_1+\commut{\hat\brst_1}{a}_0\right)
=\delta a\,,
\end{equation}
for any $a\in\qA^0$. Here, $\commut{\cdot}{\cdot}_s$ denotes the
degree $s$ component of the star-commutator.
\end{proof}

\subsection{Quantum BRST observables and non-associative star-product on $\manM$}
Given a nilpotent quantum BRST charge one can consider the 
cohomology group of its adjoint action $\hat
Q=\frac{i}{\hbar}\commut{\hat\brst}{\cdot}$. It follows from the
standard deformation theory and Theorem~\bref{thm:coh-class} that
any classical BRST cohomology class determines a quantum one. In
fact it also follows that
\begin{equation}
  H^n(\hat Q,\qA^0)\simeq H^n(Q,\aA^0)\tensor[[\hbar]]\,.
\end{equation}

It is nevertheless useful to explicitly construct representatives
of the quantum BRST cohomology classes. Similarly to the classical
case this is achieved by finding a lift of functions of $x^i,\cC^A$
to BRST invariant elements of $\qA^0$. We have the following
\begin{prop}
\label{prop:ext-q}
  For any $f_0=f_0(x,\cC,\hbar)$ there exists
  $f\in\qA^0$ such that
\begin{equation}\label{eq:q-obs}
  \commut{\brst}{f}=0\,,\qquad \gh{f}=\gh{f_0}\,,\qquad f|_{Y=\cP=0}=f_0\,.
\end{equation}
If in addition $f$ is such that $\delta^* (f-f_0)=0$ and
$\gh{f}\geq 0$ then $f$ is a unique quantum BRST-invariant
extension of $f_0$. Moreover, if $f$ and $\tilde f$ both
satisfy~\eqref{eq:q-obs} with the same $f_0$ then $f-\tilde
f=\commut{\hat\brst}{h}\,$ for some $h \in \hat\aA^0$.
\end{prop}
\begin{proof}
  The proof is standard once degree is prescribed. That equation holds
  to lowest order follows from $\delta f_0=0$.
\end{proof}
If $f,g\in \qA^0$ are unique BRST invariant extensions of functions
$f_0(x)$ and $g_0(x)$ determined by Proposition~\bref{prop:ext-q} then
one can define a bilinear operation
\begin{equation}
  f_0\star_D g_0=(f\star g)|_{Y=\cP=0}\,.
\end{equation}
This operation is not an associative product in general. However, it
determines the associative star-product on $\Sigma$. Indeed, BRST
cohomology can be identified with functions on $\Sigma$ while quantum
multiplication in $\qA^0$ determines a quantum multiplication in the
cohomology. By choosing different lifts from functions on $\manM$ to
$\qA^0$ one can describe different extensions of the associative star
products on $\Sigma$ to in general non-associative product on $\manM$.

\vspace{3mm}

As a final remark, we comment on the emergence of weak Poisson
brackets and weak star-products in the context of constrained systems.
In dynamics, and especially in what concerns the deformations and
quantization of classical dynamical systems, the Poisson geometry is
stereotypically considered the most fundamental structure of the
theory. But whenever a constrained or a gauge system is considered
that does not allow explicitly solving constraints nor taking the
quotient over the gauge symmetry, the dynamics, as such, does not
require a Poisson algebra to exist for all functions on the entire
phase-space manifold.  Only the space of physical quantities has to
carry a Poisson structure, and hence the geometry of the entire
manifold turns out to have a weaker structure than the Poisson one. As
we have seen, this is the case with non-scalar second-class
constraints. The BRST theory was originally worked out as a tool for
quantizing systems with gauge symmetries defined by weakly integrable
distributions. Now, as is seen, the idea of BRST cohomology allows one
to quantize systems whose Poisson algebra is also weak.

\subsubsection*{Acknowledgments}
This work was supported by the RFBR Grant 02-01-00930 and by the grant
INTAS 00-262. The work of IB and MG was also supported by the Grant
LSS-1578.2003.2.

\providecommand{\href}[2]{#2}\begingroup\raggedright\endgroup

\begin{thebibliography}{10}

\bibitem{Batalin:1977pb}
I.~A. Batalin and G.~A. Vilkovisky, ``Relativistic {S} matrix of dynamical
  systems with boson and fermion constraints,'' {\em Phys. Lett.} {\bf B69}
  (1977)
309--312.

\bibitem{deWit:1978cd}
B.~de~Wit and J.~W. van Holten, ``Covariant quantization of gauge theories with
  open gauge algebra,'' {\em Phys. Lett.} {\bf B79} (1978)
389.

\bibitem{Fradkin:1978xi}
E.~S. Fradkin and T.~E. Fradkina, ``Quantization of relativistic systems with
  boson and fermion first and second class constraints,'' {\em Phys. Lett.}
  {\bf B72} (1978)
343.

\bibitem{Batalin:1981jr}
I.~A. Batalin and G.~A. Vilkovisky, ``Gauge algebra and quantization,'' {\em
  Phys. Lett.} {\bf B102} (1981)
27--31.

\bibitem{BFF}
I.~A. Batalin, E.~S. Fradkin, and T.~E. Fradkina, ``Generalized canonical
  quantization of dynamical systems with constraints and curved phase space,''
  {\em Nucl. Phys.} {\bf B332} (1990) 723.

\bibitem{Fedosov-book}
B.~Fedosov, ``Deformation quantization and index theory,''. Berlin, Germany:
  Akademie-Verl. (1996) 325 p. (Mathematical topics: 9).

\bibitem{GL}
M.~A. Grigoriev and S.~L. Lyakhovich, ``Fedosov deformation quantization as a
  {BRST} theory,'' {\em Commun. Math. Phys.} {\bf 218} (2001) 437--457,
\href{http://www.arXiv.org/abs/hep-th/0003114}{{\tt hep-th/0003114}}.

\bibitem{BGL}
I.~A. Batalin, M.~A. Grigoriev, and S.~L. Lyakhovich, ``Star product for second
  class constraint systems from a {BRST} theory,'' {\em Theor. Math. Phys.}
  {\bf 128} (2001) 1109--1139,
\href{http://www.arXiv.org/abs/hep-th/0101089}{{\tt hep-th/0101089}}.

\bibitem{Cattaneo:2003dp}
A.~S. Cattaneo and G.~Felder, ``Coisotropic submanifolds in {P}oisson geometry
  and branes in the {P}oisson sigma model,'' {\em Lett. Math. Phys.} {\bf 69}
  (2004) 157--175,
\href{http://www.arXiv.org/abs/math.qa/0309180}{{\tt math.qa/0309180}}.

\bibitem{Lyakhovich:2004kr}
S.~L. Lyakhovich and A.~A. Sharapov, ``Characteristic classes of gauge
  systems,'' {\em Nucl. Phys.} {\bf B703} (2004) 419--453,
\href{http://www.arXiv.org/abs/hep-th/0407113}{{\tt hep-th/0407113}}.

\bibitem{Batalin:1992}
I.~A. Batalin and I.~V. Tyutin, ``An infinite algebra of quantum {D}irac
  brackets,'' {\em Nucl. Phys.} {\bf B381} (1992) 619--640.

\bibitem{Batalin:2001fh}
I.~Batalin and R.~Marnelius, ``Generalized {P}oisson sigma models,'' {\em Phys.
  Lett.} {\bf B512} (2001) 225--229,
\href{http://www.arXiv.org/abs/hep-th/0105190}{{\tt hep-th/0105190}}.

\bibitem{Deriglazov:1995ec}
A.~A. Deriglazov, A.~V. Galajinsky, and S.~L. Lyakhovich, ``Weak {D}irac
  bracket construction and the superparticle covariant quantization problem,''
  {\em Nucl. Phys.} {\bf B473} (1996) 245--266,
\href{http://www.arXiv.org/abs/hep-th/9512036}{{\tt hep-th/9512036}}.

\bibitem{Lyakhovich:2004xd}
S.~L. Lyakhovich and A.~A. Sharapov, ``{BRST} theory without {H}amiltonian and
  {L}agrangian,''
\href{http://www.arXiv.org/abs/hep-th/0411247}{{\tt hep-th/0411247}}.

\bibitem{Fedosov:1994}
B.~Fedosov, ``A simple geometrical construction of deformation quantization,''
  {\em J. Diff. Geom.} {\bf 40} (1994) 213--238.

\bibitem{Rothstein:1990bj}
M.~Rothstein, ``The structure of supersymplectic supermanifolds,'' {\em Lect.
  Notes Phys.} {\bf 375} (1991). (Springer-Verlag, Berlin).

\bibitem{Batalin:1991jm}
I.~A. Batalin and I.~V. Tyutin, ``Existence theorem for the effective gauge
  algebra in the generalized canonical formalism with abelian conversion of
  second class constraints,'' {\em Int. J. Mod. Phys.} {\bf A6} (1991) 3255.

\bibitem{Henneaux:1985kr}
M.~Henneaux, ``Hamiltonian form of the path integral for theories with a gauge
  freedom,'' {\em Phys. Rept.} {\bf 126} (1985) 1.

\bibitem{BGST}
G.~Barnich, M.~Grigoriev, A.~Semikhatov, and I.~Tipunin, ``Parent field theory
  and unfolding in {BRST} first-quantized terms,''
\href{http://www.arXiv.org/abs/hep-th/0406192}{{\tt hep-th/0406192}}.

\end{thebibliography}
\end{document}